\documentclass[copyright,creativecommons]{eptcs}
\providecommand{\event}{DCM 2013} 

\title{Using HMM in Strategic Games}
\author{Mario Benevides 
\qquad Isaque Lima \qquad Rafael Nader \qquad Pedro Rougemont
\institute{Systems and Computer Engineering Program and Computer Science Department}
\institute{ Federal University of Rio de Janeiro, Brazil}
}
\def\titlerunning{Using HMM in Strategic Games}
\def\authorrunning{M. Benevides, I. Lima, R. Nader \& P. Rougemont}

\usepackage{graphicx}
\usepackage{float}

\begin{document}
\maketitle

\begin{abstract}
In this paper we describe an approach to resolve strategic games in which players can assume different types along the game. Our goal is to infer 
which type the opponent is adopting at each moment so that we can increase the player's odds. To achieve that we use Markov games combined with 
hidden Markov model. We discuss a hypothetical example of a tennis game whose solution can be applied to any game with similar characteristics.

\end{abstract}


\newtheorem{thm}{Theorem}
\newtheorem{cor}{Corollary}
\newtheorem{lem}{Lemma}
\newtheorem{defn}{Definition}
\newtheorem{exmp}{Example}



\section{Introduction}

Game theory is widely used to model various problems in economics, the field area in which it was originated. It has been increasingly used in different applications, we can highlight their importance in political and diplomatic relations, biology, computer science, among others.
The study of game theory relies on describing, modeling and solving problems directly related to interactions between rational decision makers. We are interested in matches between two players that choose their actions in the best way, and know that each other do the same.

In this paper we propose a model which maps opponent players behavior as a set of states (types),
each state having a pre-defined payoff table, in order to infer opponents next move. We address the
problem in which the states cannot be directly observed, but instead, they can be estimated from the
observations of the players actions.

The rest of this paper is organized as follows. In section 2 we show related works and the motivation to our study. In section 3, we present the necessary background in Hidden Markov Model. In section 4 we introduce our model and in section 5 we illustrate it using a tennis game example, whose solution can be applied to any game with similar characteristics. In appendix A, we provide a Game Theory tool kit for the reader not familiar with this subject.

\section{Related Work}\label{RW}

In this paper we try to solve a particular case of a problem known as Repeated Game with Incomplete Information, first introduced by Aumann and Maschler in 1960 \cite{aumann95}, described below:

The game G is a Repeated Game where in the first round the state of Nature is chosen by a probability p and only player 1 knows this state. Player 2 knows the possible states, but doesn’t know the actual state. After each round, both players know the action of each one, and play again.


In \cite{renault06} it is studied  Markov Chain Games with Lack of Information. In this game, instead of a probability associated to an initial state of Nature, we have a Markov Chain Transition Probability Distribution that changes the state through  time. Both players know the action of each other at each round, but only one player knows the actual and past states and the payoff associated with those actions. The other player knows the Transition Probability Distribution. This game is a particular case of Markov Chain Game or Stochastic Game \cite{markovgames}, where both players know the actual state (or the last state and the transition probability distribution).

\cite{renault06}  presents some properties and solutions for this class of games, using the recurrence property of a Markov Chain and the Belief Matrix B.

Our problem is a particular case of Renault's game, where the lack of information is worse and the unaware player does not know the Transition Probability Distribution of the Markov Chain Game.  With that, we can describe formally our problem as below:

The game G is a Repeated Game between two players where the Player 1 has a type that changes through the time. Each type has a probability distribution to other types that is independent of past states and actions. Each type introduces a different game with different payoffs. Player 1 knows his actual type, player 2 knows the types that player 1 can be, but does not know the actual type nor the transition between types. Both players know the action of each round.

As mentioned before, our game is a particular case of a Markov Chain Game, so the transition between types of player 1 follows the Markov Property. But as the player 2 doesn’t know the state of player 1 we can’t solve this as a Markov Chain Game (or Stochastic Game), and as the player 2 does not know the transition between the types we can not use the Markov Chain Game with Lack of Information.

To solve this problem we propose a model to this game and a solution that involves Hidden Markov Model. We compare our results with other ways to solve a problem like that.

\section{Hidden Markov Model (HMM)}\label{HMM}

A Markov model is a stochastic model composed by a set of states to describe a sequence of events. It has the Markov property, which means that the process is memoryless, i.e., the next state depends only on the present state. A hidden Markov model \cite{rabiner85,rabiner89} is a Markov model in which the states are partially observable.

            In a hidden Markov model we have a set of hidden states called hidden chain, that is exactly like a Markov chain, and we have a set of observable states. Each state in the hidden chain is associated to the observable states with a probability. We do not know the hidden chain, neither the actual state, but we know the actual observation. The idea is that after a series of observations we can get information about the hidden chain.

\begin{figure}[htb]
\centering
\begin{minipage}[b]{0.5\linewidth}
\includegraphics[bb=1.0in 0.6in 8.5in 8.5in, scale=0.50]{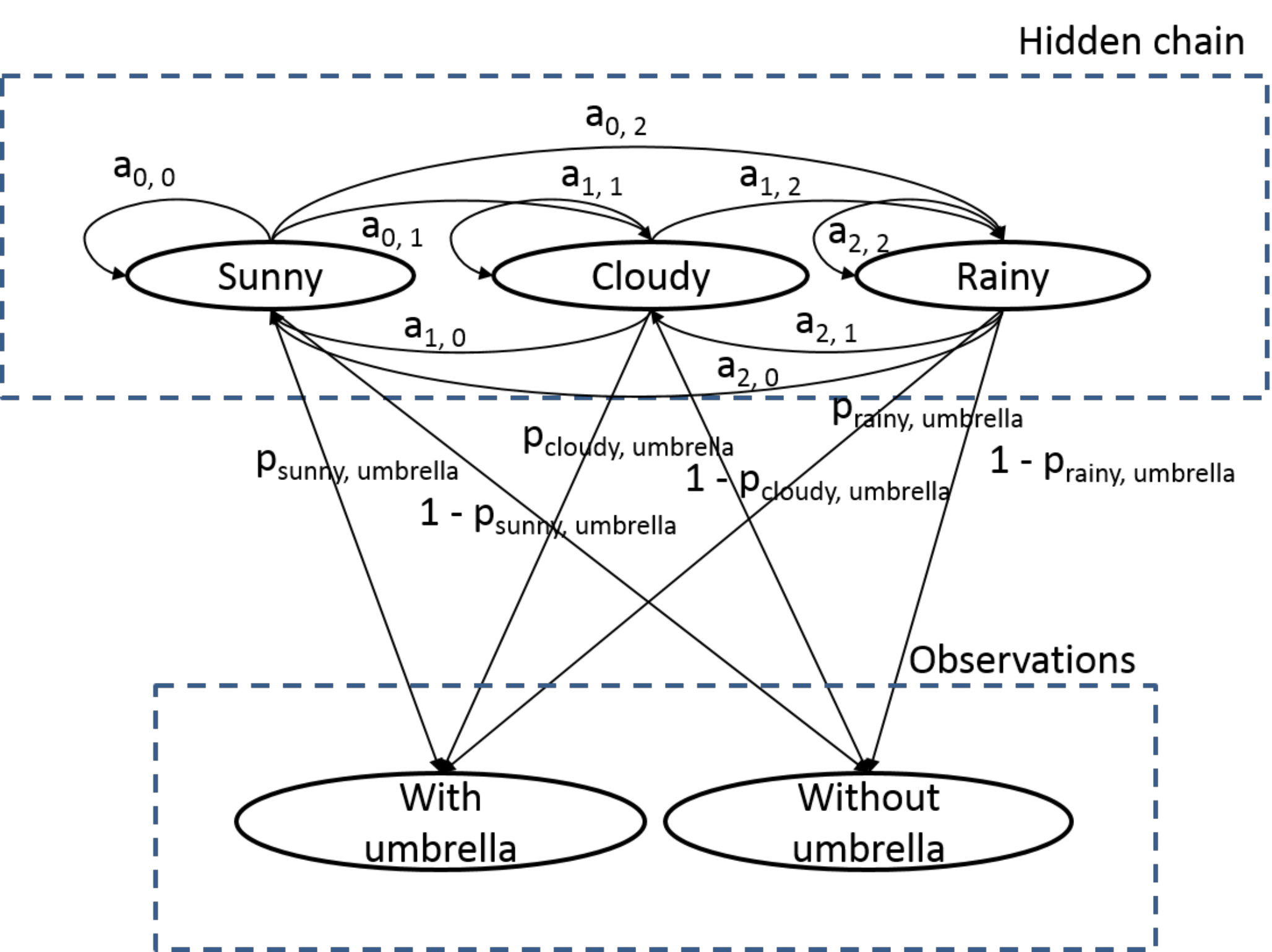}
\caption{HMM Example}
\label{temp} \end{minipage}
\end{figure}

An example to illustrate the use of hidden Markov model is a three state weather problem (see figure \ref{temp}). We are interested in help an office worker that is locked in the office for several days to know about the weather outside. Here we have three weather possibilities: rainy, cloudy and sunny, that changes through the time depending only on the actual state. The worker observes people coming to work with an umbrella sometimes and associates a probability to each weather based on that. With a series of umbrella observations, we would like to know about the weather, the hidden state.

\begin{defn} A  Hidden Markov Model  $\Lambda = \langle A,B, \Pi \rangle$ consists of

\begin{itemize}
\item a finite set of states $Q = \{q_1, ..., q_n\}$;

\item  a  transition array $A$, storing the probability of the current state be $q_i$ given that the
 previous state was $q_j$;

\item  a finite set of observation states $O = \{o_1, ..., o_k\}$;

\item a  transition array $B$, storing the probability of the observation state $o_i$ be produced
from state  $q_j$;

\item a $1 \times n$ initial array $\Pi$, storing the initial probability of the model be in state $q_i$.

\end{itemize}
 
\end{defn}

\begin{figure}[htb]
\centering
\begin{minipage}[b]{0.5\linewidth}
\includegraphics[bb=1.0in 0.6in 8.5in 8.5in, scale=0.50]{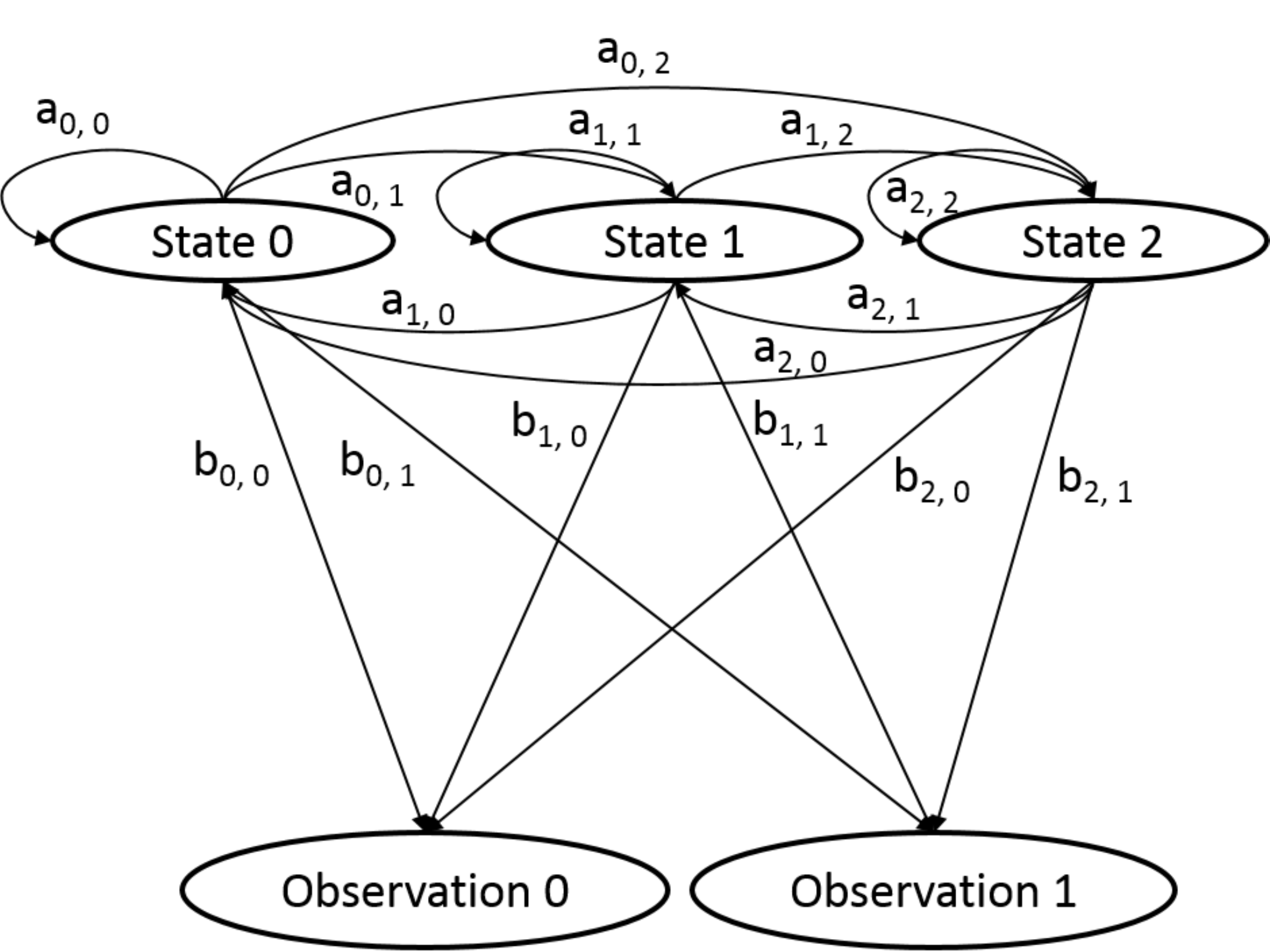}
\caption{HMM Definition}
\label{hmmdf} \end{minipage}
\end{figure}

Based on a given set of observations “$y_i$”, and a predefined set of transitions “$b_{ij}$”, 
the Hidden Markov Model’s framework allows us to estimate the hidden transitions, labeled “$a_{ij}$” in the figure \ref{hmmdf}.

\section{Hidden Markov Game (HMG)}\label{HMG}

In this section, we describe the Hidden Markov Game (HMG), which is inspired by the works discussed in section \ref{RW}. We first define it for
many player and then restrict it to two players.

 \begin{defn} A Hidden Markov Game $G$ consists of

\begin{itemize}
\item A finite set of players $P = \{1, ..., I \}$;
\item Strategy sets $S_1, S_2, ... S_I$, one for each player;
\item Type sets $T_1, T_2, ... T_I$, one for each player;

\item A transition functions ${\cal T}_i ~:~ T_i \times S_1 \times ... \times S_I \mapsto {\cal PD}(T_i)$, one for each player;

\item A finite set of observation state variables ${\bf O} = \{{\cal O}_1, ..., {\cal O}_k\}$;


\item Payoff functions $u_i: S_1 \times S_2 \times ... \times S_I  \times T_i \mapsto \Re$, one for each player;
\item Probability Distribution $\pi_{i,j}$, representing the player $i$ prior belief about the type of his opponent j, for each player j.
\end{itemize}

Where  ${\cal PD}(T_i)$ is a probability distribution over $T_i$.

\end{defn}

We are interested in a particular case of the HMG where we have two players and players 1 does not know his transitions function ${\cal T}_i$. 
And the set of observation state variables ${\bf O} = S_2$

We are interested in a particular case of the HMG:

\begin{itemize}

\item        two players

\item        player 1 does not know the opponent transition function

\item         The observation state variables O for player 1 are equal to the set of strategies of the opponent

\end{itemize}

\bigskip

 {\bf Problem}: Given a sequence of observations on time of the  observation state variables $O^{t_0}, ...., O^{t_m}$, we want to know the probability, for player $i$, that  $O^{t_{m + 1}} = {\cal O}_j$, for $1\leq j \leq k$.

In fact, a Hidden Markov Game can be seem as a Markov Game where a player does not know the probability distributions of the transitions between types. Instead, he knows a sequence of observation on time of the observable behavior of each opponent.

Due to this particularity, we cannot use the standard Markov Game tools to solve the game. The aim of our model is to provide a solution for the game partitioning the problem. In order to achieve that we infer the Markov chain at each turn of the game and then play in accordance to this Markov Game. In fact, if we had a one turn game, then we would have a Bayesian Game and we could calculate the Bayesian Nash Equilibrium and play according to it. This  is what has been done in \cite{waghabi}.

The inference of the transitions of the Markov chain is accomplished using a Hidden Markov Model HMM. The probability distribution associating each state of the Markov chain to the observable states are given by the Mixed Equilibrium of each matrix in each type. A Baum Welch Algorithm is used to infer the HMM.

\bigskip

\noindent {\bf Our Solution}: Given a Hidden Markov Game $G$ and a sequence of observations on time of the  observation state variables $O^{t_0}, ...., O^{t_m}$, we want to calculate the probability, for player $i$, that $O^{t_{m + 1}} = {\cal O}_j$, for $1\leq j \leq k$. This is accomplished following the steps below:

\begin{enumerate}

\item Represent the Hidden Markov Game $G$ as a Hidden Markov Model each for player $i$ as follows:

\begin{enumerate}

\item {\it the finite set of states $Q = \{q_1, ..., q_n\}$} is the set the set $T_i \times S_1 \times ... \times S_I$;

\item   {\it the transition array $A$, storing the probability of the current state be $q_j$ given that the
 previous state was $q_l$} is what we want to infer;

\item   {\it the finite set of observation states $O = \{o_1, ..., o_k\}$} is the set of observation state variables ${\bf O} = \{{\cal O}_1, ..., {\cal O}_k\}$;

\item {\it the transition array $B$, storing the probability of the observation state ${\cal O}_h$ be produced
from state  $q_j = <T_j, s_1, ... , s_I>$} is obtained calculating the probability of player $i$ using strategy $< s_1, ... , s_I>$ in the Mixed Nash Equilibrium of game $T_j$ and adding the probabilities, for each profile $\vec{s}^{-i} \in f_i({\cal O}_h)$;

\item {\it the $1 \times n$ initial array $\Pi$, storing the initial probability of the model be in state $q_j$} is the $\pi_{i,j}$.

\end{enumerate}

\item Use the Baum Welsh algorithm to infer the matrix $A$;

\item Solve the underline Markov Game and find the must probable type each opponent is playing;

\item If it is one move game then choose the observable state with the greatest probability. Else, play according to the Mixed Nash Equilibrium.

\end{enumerate}

\section{Application and Test}\label{A}

In order to illustrate our frame work we present an example of a tennis game. In this example, we have two tennis players, and we are interested in the server versus receiver situation. Here, we can see the possible set of strategies for each one as:
\begin{itemize}
\item          Server can try to serve a central ball (center) or an open one (open).

\item         Receiver will try to be prepared to one of those actions.
\end{itemize}

We have different server types: aggressive, moderate or defensive, what means that the way he chooses the strategy will vary. For each type, the player evaluate each strategy in a different way. The player's type changes through the time as a Markov chain.

We are interested in helping the receiver to choose which action to take, but we need to know about opponent actual type.

We assume that the transitions between the hidden states and the observations are fixed, i.e., they are previously computed using 
the payoff matrix of each profile. We do that to reduce the number of loose variables.
We compare ours results with the ones obtained by Bayesian game using the same trained HMM to compute the Bayesian output. We also compare our results with some more naive approaches like Tit-for-Tat (that always repeat the last action made by the opponent ), Random Choice  and More Frequently (choose the action that is more frequently used by the opponent).

We used the following payoff matrixes, with two players: player 1 (column) and player 2 (row).

\begin{table}[h]
 \centering
\footnotesize
\begin{tabular}{c|c|c|c|}
	&	Open		&	Center\\ \hline
Open	&	0.65,0.35	&	0.89,0.11 \\ \hline 
Center	&	0.98,0.02	&	0.15,0.85 \\ \hline

\end{tabular}

\caption{Payoff Matrix (Aggressive Profile)}
\label{agg_game}
\end{table}

\begin{table}[h]
 \centering
\footnotesize
\begin{tabular}{c|c|c|c|}
	&	Open		&	Center\\ \hline
Open	&	0.15,0.85	&	0.80,0.20 \\ \hline 
Center	&	0.90,0.10	&	0.15,0.85 \\ \hline

\end{tabular}

\caption{Payoff Matrix (Moderate Profile)}
\label{mod_game}
\end{table}

\begin{table}[h]
 \centering
\footnotesize
\begin{tabular}{c|c|c|c|}
	&	Open		&	Center\\ \hline
Open	&	0.10,0.90	&	0.55,0.45 \\ \hline 
Center	&	0.85,0.15	&	0.05,0.95 \\ \hline

\end{tabular}

\caption{Payoff Matrix (Defensive Profile)}
\label{def_game}
\end{table}

We  compute the mixed strategy to each profile, and consequentially compute 
the values of B. 

Next we present some scenarios that we use.

\subsection{Scenarios}

In this section we present some of the scenarios that we use. 

The original HMM is only used to generate the observations of the game.
We generate 10.000 observations, for each scenario, and after every 200 observations we compute the result of the game.

We used the metric proposed in \cite{rabiner89} to compare the similarity of two Markov models.

\subsubsection{Scenario 1 – Aggressive Player}

In order to illustrate the behavior of an aggressive player we model a HMM that has more probability to stay in the aggressive state.

\begin{figure}[H]
\centering
\includegraphics[scale=0.50]{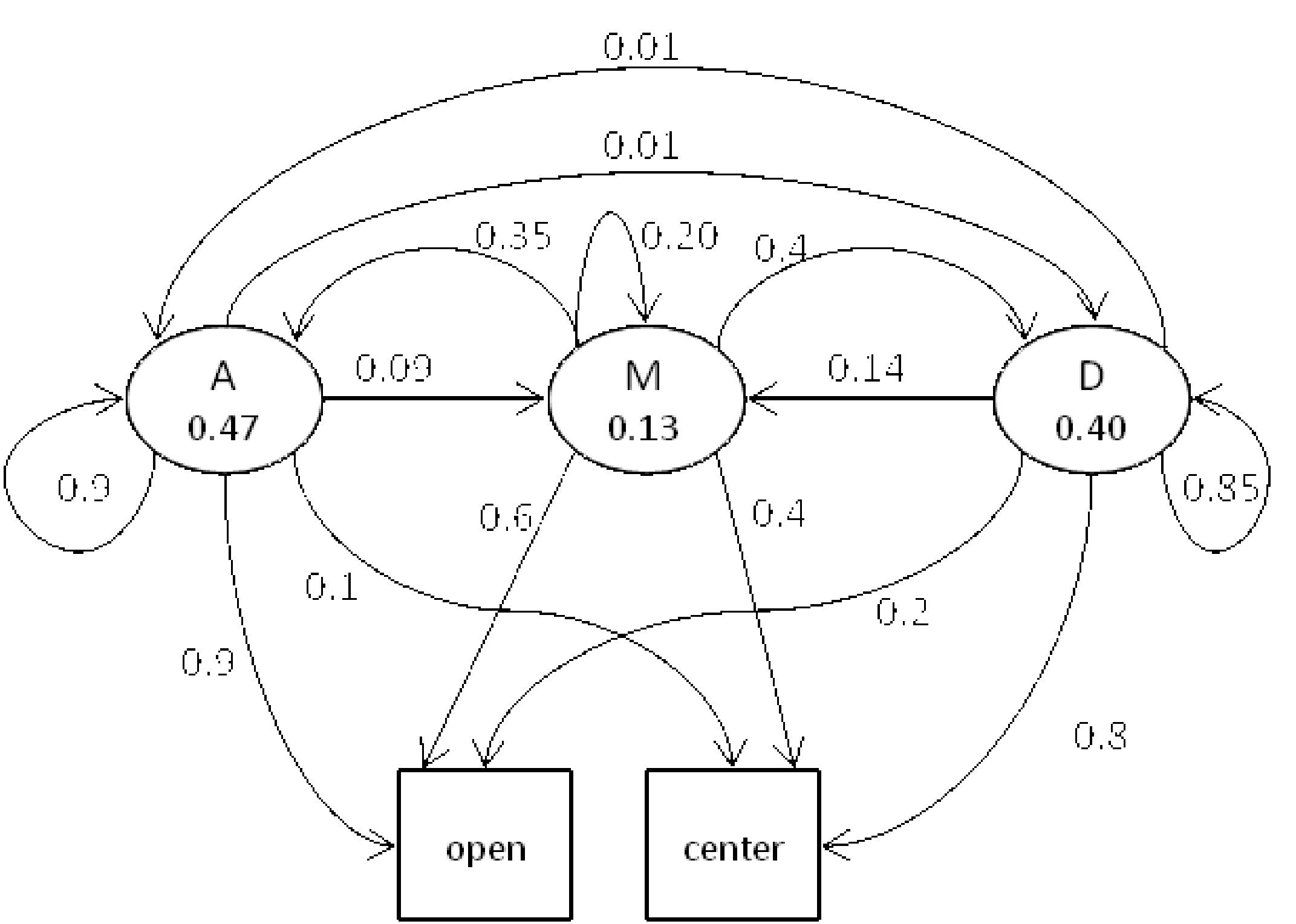}
\caption{Original HMM (Aggressive Player)}
\end{figure}

\begin{figure}[H]
\centering
\includegraphics[scale=0.50]{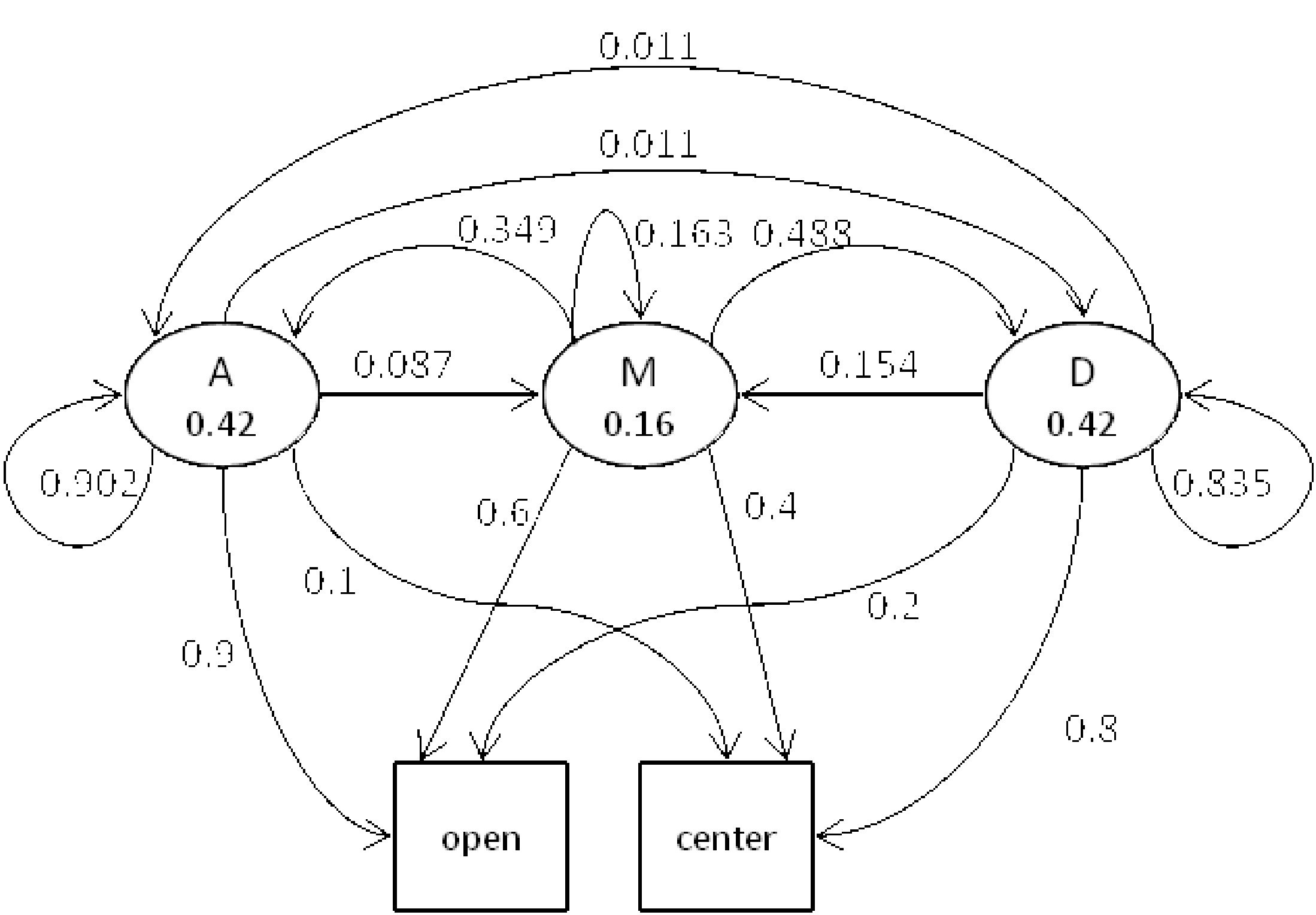}
\caption{ Trained HMM ( Aggressive Player)}
\end{figure}

The difference between the original HMM and de trained HMM was 0,0033, i.e, the two HMM are very close, as we can empiric see in the picture below.
As we can see on the table below, the proposed algorithm increase the player odds.

\begin{table}[H]
 \centering
\footnotesize
\begin{tabular}{c|c|c|c|c|c|c|}
	&	Proposed Model		&	Bayesian Game	&	 Random	&	More Frequently	&	Tit-for-Tat \\ \hline
hit rate	&	0,78	&	0,58	&	0,496		&	0,582		&	0,532 \\ \hline 

\end{tabular}
\caption{Agressive Scenario Hit Rate}
\label{aggschitrate}
\end{table}

\begin{figure}[H]
\centering\begin{minipage}[b]{0.5\linewidth}
\includegraphics[bb=1.0in 0.6in 8.5in 8.5in, scale=0.60]{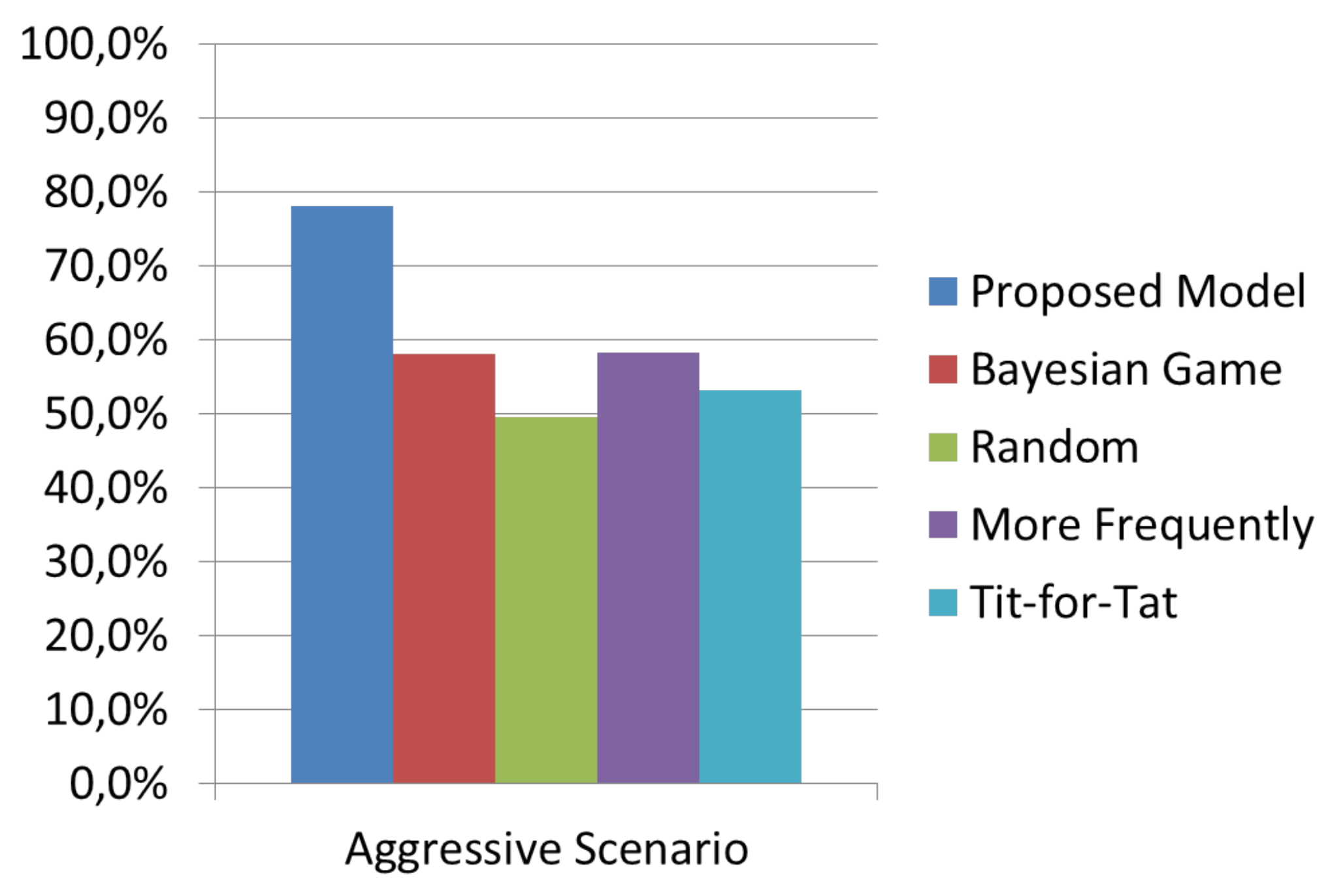}
\caption{Agressive Scenario Graphic}\end{minipage}
\end{figure}

\subsubsection{Scenario 2 – Defensive Player}

In order to illustrate the behavior of a defensive player we model a HMM that has more probability to stay in the defensive state.

\begin{figure}[H]
\centering
\includegraphics[scale=0.50]{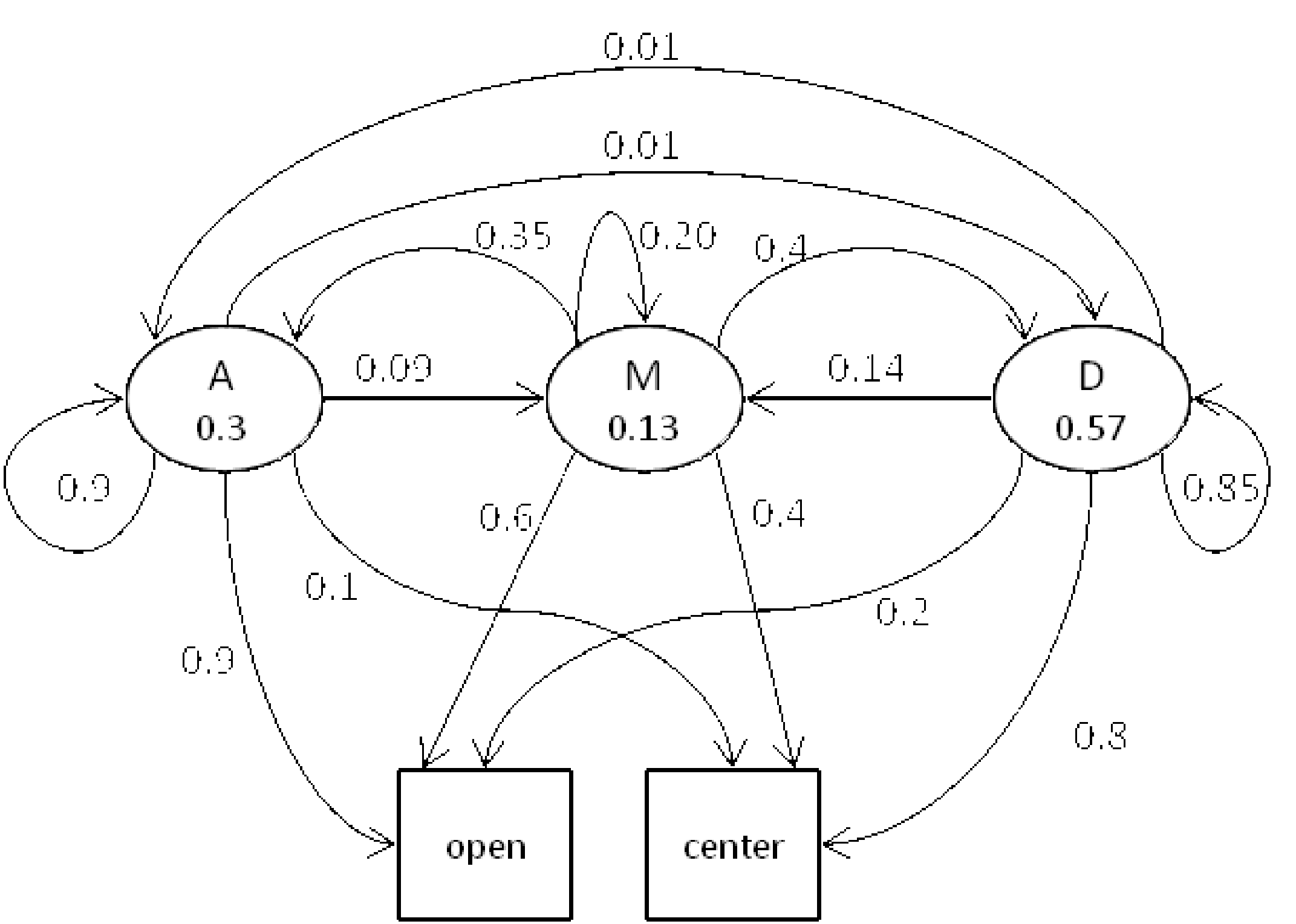}
\caption{Original HMM (Defensive Player)}
\end{figure}

\begin{figure}[H]
\centering
\includegraphics[scale=0.50]{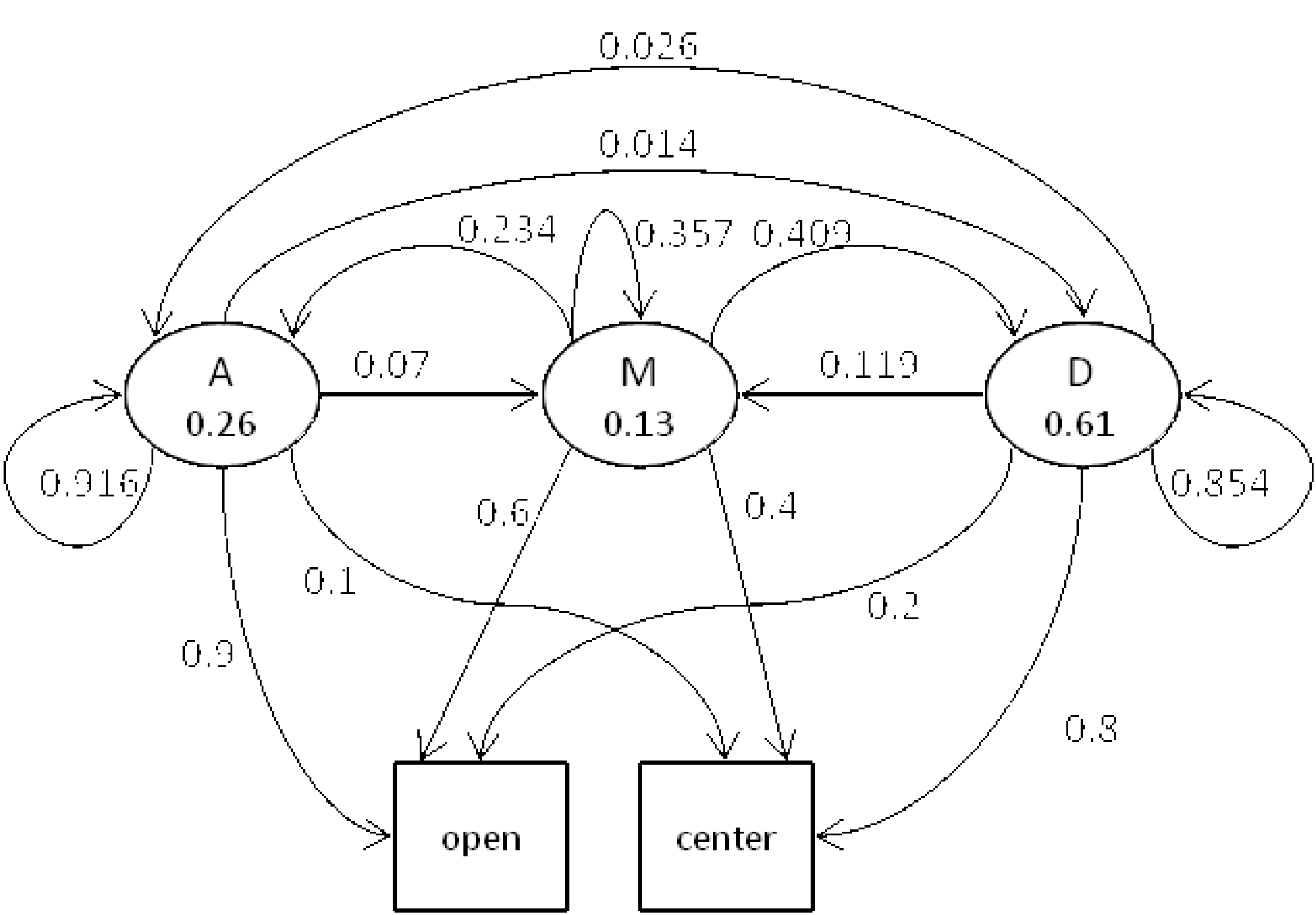}
\caption{Trained HMM (Defensive Player)}
\end{figure}


Once again the proposed algorithm increase the player odds and the trained HMM is very close to the original HMM.

\begin{table}[H]
 \centering
\footnotesize
\begin{tabular}{c|c|c|c|c|c|c|}
	&	Proposed Model		&	Bayesian Game	&	 Random	&	More Frequently	&	Tit-for-Tat \\ \hline
hit rate	&	0,71	&	0,50	&	0,492		&	0,53		&	0,498 \\ \hline 

\end{tabular}
\caption{Defensive Scenario Hit Rate}
\label{defschitrate}
\end{table}

\begin{figure}[H]
\centering \begin{minipage}[b]{0.5\linewidth}
\includegraphics[bb=1.0in 0.6in 8.5in 8.5in, scale=0.60]{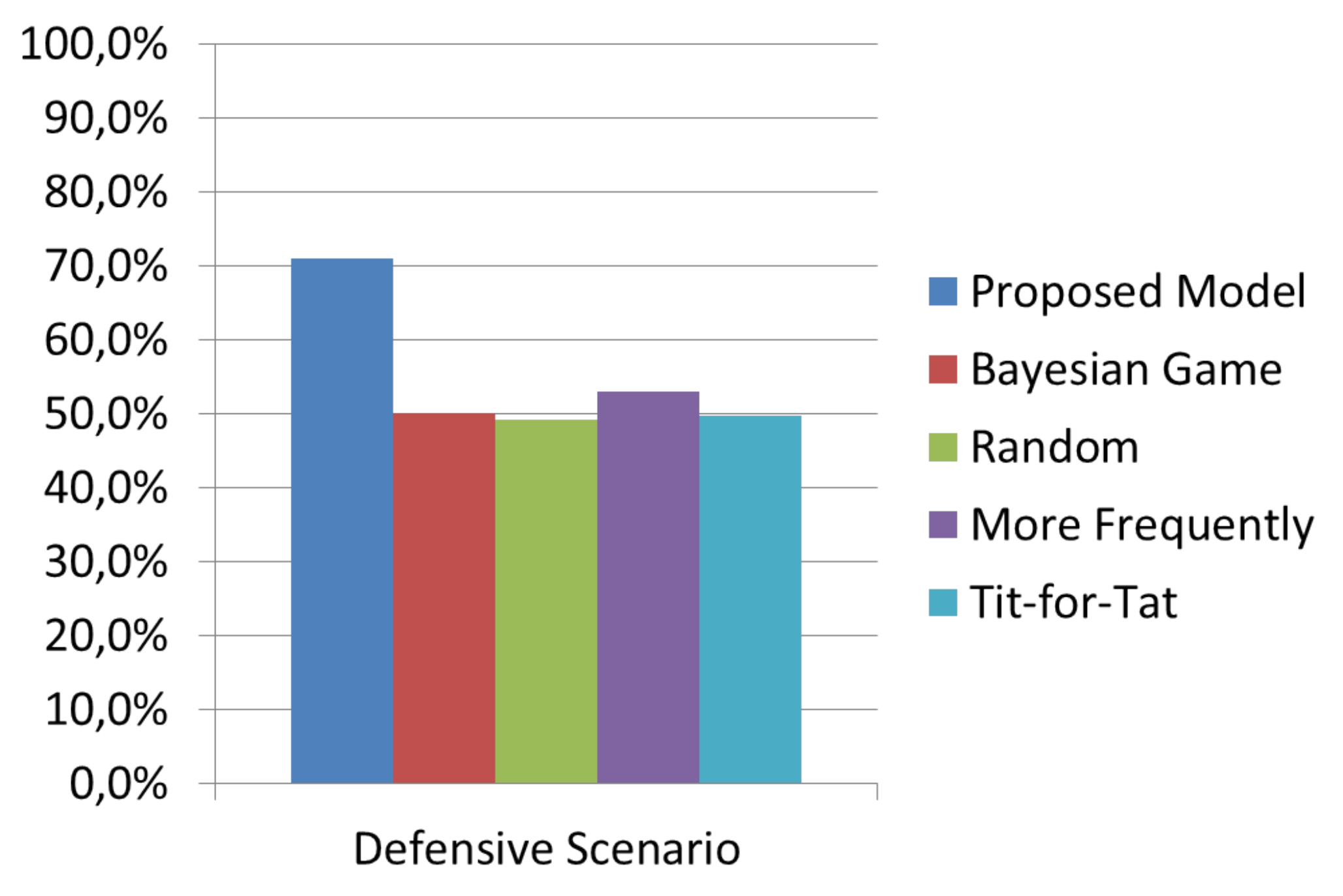}
\caption{Defensive Player Graphic}\end{minipage}
\end{figure}

The gain in these two scenarios are due to the fact that the proposed algorithm take into account the transitions between hidden states, so we have more information that help us to choose a better strategy.

\section{Conclusions}

In this work we have introduced a novel class of games called Hidden Markov Games, which can be thought as Markov Game where a player does not know the probability distributions of the transitions between types. Instead, he knows a sequence of observation on time of the observable behavior of each opponent. 
 We propose a solution to our game representing it  as a Hidden Markov Model, We use Baum Welsh algorithm to infer the probability distributions of the transitions between types. Finally, we solve the underline Markov Game.

In order to illustrate our approach, we present a tennis game example and solve it using our method. The experimental results indicates that our solution is quite good.

In this work we present a solution for a special case of  Markov Chain Games or Stochastic Games with the lack of information  and where the unaware player does not know the Transition Probability Distribution of the Markov Chain Game.  We provide a solution based on Hidden Markov Models which is computational and quite efficient. The drawback of our approach is the fact that we use the Nash Equilibrium of each type in a ad hoc way.

As future work we would like to prove that playing with the Nash equilibrium of actual type is an optimal choice or at least an equilibrium in the repeated game. Another problem to solve is to show that the choice of the uninformed player in playing with an equilibrium of the  infered type is an optimal choice or at least a Nash equilibrium of whole game.

\appendix 


\section{Game Theory Tool Kit}\label{ap_A}

In this section, we present  the necessary background on Game theory. First we introduce the concepts of Normal Form Strategic Game and Pure and Mixed Nash Equilibrium. Finally, we define Bayesian  Games and Markov Games.

\begin{defn} A Normal (Strategic) Form game $G$ consists of

\begin{itemize}
\item A finite set of players $P = \{1, ..., I \}$;
\item Strategy sets $S_1, S_2, ... S_I$, one for each player;
\item Payoff functions $u_i: S_1 \times S_2 \times ... \times S_I \mapsto \Re$, one for each player.
\end{itemize}

A {\it strategy profile} is a tuple $s= s_1,s_2,...,s_I$ such that $s \in {\cal S}$, where
${\cal S} = S_1 \times S_2 \times ... \times S_I$. We denote $s_{-i}$ as the profile obtained from $s$ removing $s_i$, i.e.,
$s_{-i}= s_1,s_2,...,s_{i-1}, s_{i+1}, ..., s_I$
 
\end{defn}

The following game is an example of normal (strategic) form game with two players Rose (Row) and Collin (Column) and both have two strategies $s_1$ and $s_2$.

\begin{table}[h]
\begin{center}

\begin{tabular}{c|c|c|c|}
		&	$s_1$	&	$s_2$\\ \hline
$s_1$	&	3,3	&	2,5 \\ \hline 
$s_2$	&	5,2	&	1,1 \\ \hline

\end{tabular}

\end{center}

\caption{Normal Form Strategic Game}
\label{mix_game}
\end{table}

\begin{defn} A strategy profile $s^*$ is a pure strategy Nash equilibrium of $G$ if and only if

$$u_i(s^*) \geq u_i(s_i,s_{-i})$$ 

for all players $i$ and all strategy $s_i \in S_i$, where $u$ is the payoff function.

\end{defn}

Intuitively, a strategy profile is a Nash Equilibrium if for each player, he cannot improve his payoff changing his strategy alone.

 The game presented in table \ref{mix_game} has two Nash Equilibrium ($s_1,s_2$) and ($s_2,s_1$). Which one they should play? A possible answer could be to assign probabilities to the strategies, i.e., Rose could play $s_1$ with probability $p$ and $s_2$ with $1-p$, and Collin could play $s_1$ with probability $q$ and $s_2$ with $1-q$. This game has a mixed Nash equilibrium that is $p=1/3$ and $q=1/3$. We can calculate the expected payoff for both player and check that they cannot improve their payoff changing their strategy alone, i.e., this a Nash Equilibrium.

 Other class of static games are strategic games of incomplete information also called Bayesian Games \cite{osborne,gibbons}. The intuition behind this kind of game is that in some situations the player knows his payoff function but is not sure about his opponents function. He knows that his opponent is playing according to one type in a finite set of types with some probability.
 
 \begin{defn} A Static Bayesian game $G$ consists of

\begin{itemize}
\item A finite set of players $P = \{1, ..., I \}$;
\item Strategy sets $S_1, S_2, ... S_I$, one for each player;
\item Type sets $T_1, T_2, ... T_I$, one for each player;
\item Payoff functions $u_i: S_1 \times S_2 \times ... \times S_I  \times T_i \mapsto \Re$, one for each player.
\item Probability Distribution $P(t_{-i} \mid t_i)$ , denoting the player $i$ belief about his opponents types $t_{-i}$, given that his type is $t_i$.
\item Probability Distribution $\pi_i$, representing the player $i$ prior belief about the types of his opponents.
\end{itemize}
 
\end{defn}

Markov Games  \cite{markovgames,owen,littman} can be though as a natural extension of Bayesian games, where we have transitions between types and a probability distributions over these transitions. 

A Markov game can be defined as follows

 \begin{defn} A Markov game $G$ consists of

\begin{itemize}
\item A finite set of players $P = \{1, ..., I \}$;
\item Strategy sets $S_1, S_2, ... S_I$, one for each player;
\item Type sets $T_1, T_2, ... T_I$, one for each player;
\item A transition functions ${\cal T}_i ~:~ T_i \times S_1 \times ... \times S_I \mapsto {\cal PD}(T_i)$, one for each player;
\item Payoff functions $u_i: S_1 \times S_2 \times ... \times S_I  \times T_i \mapsto \Re$, one for each player;
\item Probability Distribution $\pi_i$, representing the player $i$ prior belief about the types of his opponents.
\end{itemize}

Where  ${\cal PD}(T_i)$ is a probability distribution over $T_i$.
 
\end{defn}

\bibliographystyle{eptcs}
\bibliography{ref}

\begin{thebibliography}{10}
\providecommand{\bibitemdeclare}[2]{}
\providecommand{\surnamestart}{}
\providecommand{\surnameend}{}
\providecommand{\urlprefix}{Available at }
\providecommand{\url}[1]{\texttt{#1}}
\providecommand{\href}[2]{\texttt{#2}}
\providecommand{\urlalt}[2]{\href{#1}{#2}}
\providecommand{\doi}[1]{doi:\urlalt{http://dx.doi.org/#1}{#1}}
\providecommand{\bibinfo}[2]{#2}

\bibitemdeclare{book}{aumann95}
\bibitem{aumann95}
\bibinfo{author}{R.~J. \surnamestart Aumann\surnameend} \&
  \bibinfo{author}{M.~B. \surnamestart Maschler\surnameend}
  (\bibinfo{year}{1995}): \emph{\bibinfo{title}{Repeated Games with Incomplete
  Information}}.
\newblock \bibinfo{publisher}{MIT Press}, \bibinfo{address}{Cambridge, MA}.
\newblock \bibinfo{note}{With the collaboration of Richard E. Stearns}.

\bibitemdeclare{book}{gibbons}
\bibitem{gibbons}
\bibinfo{author}{Robert \surnamestart Gibbons\surnameend}
  (\bibinfo{year}{1992}): \emph{\bibinfo{title}{A primer in game theory}}.
\newblock \bibinfo{publisher}{Harvester Wheatsheaf}.

\bibitemdeclare{inproceedings}{littman}
\bibitem{littman}
\bibinfo{author}{Michael~L. \surnamestart Littman\surnameend}
  (\bibinfo{year}{1994}): \emph{\bibinfo{title}{Markov Games as a Framework for
  Multi-Agent Reinforcement Learning}}.
\newblock In \bibinfo{editor}{William~W. \surnamestart Cohen\surnameend} \&
  \bibinfo{editor}{Haym \surnamestart Hirsh\surnameend}, editors: {\sl
  \bibinfo{booktitle}{Proceedings of the Eleventh International Conference on
  Machine Learning}}, \bibinfo{publisher}{Morgan Kaufmann}, pp.
  \bibinfo{pages}{157--163}.

\bibitemdeclare{book}{osborne}
\bibitem{osborne}
\bibinfo{author}{M.J. \surnamestart Osborne\surnameend} \&
  \bibinfo{author}{A.~\surnamestart Rubinstein\surnameend}
  (\bibinfo{year}{1994}): \emph{\bibinfo{title}{A Course in Game Theory}}.
\newblock \bibinfo{publisher}{MIT Press}.

\bibitemdeclare{book}{owen}
\bibitem{owen}
\bibinfo{author}{M.~\surnamestart Owen\surnameend} (\bibinfo{year}{1982}):
  \emph{\bibinfo{title}{Game Theory}}.
\newblock \bibinfo{publisher}{Academic Press}.
\newblock \bibinfo{note}{Second Edition}.

\bibitemdeclare{article}{rabiner85}
\bibitem{rabiner85}
\bibinfo{author}{L.~R. \surnamestart Rabiner\surnameend}
  (\bibinfo{year}{1985}): \emph{\bibinfo{title}{A Probabilistic Distance
  Measure for Hidden Markov Models}}.
\newblock {\sl \bibinfo{journal}{AT$\&$T Tech. J.}}
  \bibinfo{volume}{64}(\bibinfo{number}{2}), pp. \bibinfo{pages}{391--408}.

\bibitemdeclare{book}{rabiner89}
\bibitem{rabiner89}
\bibinfo{author}{L.~R. \surnamestart Rabiner\surnameend}
  (\bibinfo{year}{1989}): \emph{\bibinfo{title}{A Tutorial on Hidden Markov
  Models and Selected Applic}}.
\newblock \bibinfo{publisher}{IEEE}.
\newblock \bibinfo{note}{Speech Recognition}.

\bibitemdeclare{article}{renault06}
\bibitem{renault06}
\bibinfo{author}{J{\'e}r{\^o}me \surnamestart Renault\surnameend}
  (\bibinfo{year}{2006}): \emph{\bibinfo{title}{The Value of Markov Chain Games
  with Lack of Information on One Side}}.
\newblock {\sl \bibinfo{journal}{Math. Oper. Res.}}
  \bibinfo{volume}{31}(\bibinfo{number}{3}), pp. \bibinfo{pages}{490--512}.
\newblock \urlprefix\url{http://dx.doi.org/10.1287/moor.1060.0199}.

\bibitemdeclare{mastersthesis}{waghabi}
\bibitem{waghabi}
\bibinfo{author}{E.~\surnamestart Waghabir\surnameend} (\bibinfo{year}{2009}):
  \emph{\bibinfo{title}{Applying HMM in Mixed Strategy Game}}.
\newblock Master's thesis, \bibinfo{school}{COPPE-Sistemas}.

\bibitemdeclare{book}{markovgames}
\bibitem{markovgames}
\bibinfo{author}{J.~Van~Der \surnamestart Wal\surnameend}
  (\bibinfo{year}{1981}): \emph{\bibinfo{title}{Stochastic dynamic
  programming}}.
\newblock \bibinfo{publisher}{M. Kaufmann}.
\newblock \bibinfo{note}{In Mathematical Centre Tracts, 139}.

\end{thebibliography}

\end{document}